\documentclass[10pt,aps,prl,superscriptaddress,reprint]{revtex4-1}

\usepackage{hyperref}
\hypersetup{
	pdfborder={0 0 0},
	pdfnewwindow=true,      % links in new window
	colorlinks=true,       % false: boxed links; true: colored links
	linkcolor=blue,          % color of internal links (change box color with linkbordercolor)
	citecolor=blue,        % color of links to bibliography
	filecolor=blue,      % color of file links
	urlcolor=blue,          % color of external links
}
\usepackage{graphicx} % Include figure files
\usepackage{amsmath} % use \text
\usepackage{amsfonts} % use \mathbb
\usepackage{xcolor}
\usepackage{ulem}

\usepackage{float}
\usepackage{array}
\usepackage{threeparttable} 
\usepackage{multirow}

\DeclareMathAlphabet{\mathsfsl}{OT1}{cmss}{m}{sl}

\begin{document}

% Use the \preprint command to place your local institutional report
% number in the upper righthand corner of the title page in preprint mode.
% Multiple \preprint commands are allowed.
% Use the 'preprintnumbers' class option to override journal defaults
% to display numbers if necessary
%\preprint{}

%Title of paper
\title{Dynamical correlation enhanced orbital magnetization in VI$_{3}$}

\author{Zhimou Zhou}
\affiliation{International Center for Quantum Materials, School of Physics, Peking
	University, Beijing 100871, China}

\author{Shishir Kumar Pandey}
\affiliation{International Center for Quantum Materials, School of Physics, Peking
	University, Beijing 100871, China}

\author{Ji Feng}
\email{jfeng11@pku.edu.cn}
\affiliation{International Center for Quantum Materials, School of Physics, Peking
	University, Beijing 100871, China}
\affiliation{Collaborative Innovation Center of Quantum Matter, Beijing 100871,
	China}
%\affiliation{CAS Center for Excellence in Topological Quantum Computation, University of Chinese Academy of Sciences, Beijing 100190, China}

\date{\today}
\begin{abstract}
The effect of electronic correlations on the orbital magnetization in real materials has not been explored beyond a static mean-field level. Based on the dynamical mean-field theory, the effect of electronic correlations on the orbital magnetization in layered ferromagnet VI$_3$ has been studied. A comparison drawn with the results obtained from density functional theory calculations robustly establishes the crucial role of dynamical correlations in this case. In contrast to the density functional theory that
leads to negligible orbital magnetization in VI$_3$, in dynamical mean-field approach the orbital magnetization is greatly enhanced. Further analysis show that this enhancement is mainly due to the enhanced local circulations of electrons, which can be attributed to a better description of the localization behavior of correlated electrons in VI$_3$.
The conclusion drawn in our study could be applicable to a wide range of layered materials in this class.
\\

\end{abstract}
\maketitle

% insert suggested PACS numbers in braces on next line
%\pacs{}
% insert suggested keywords - APS authors don't need to do this
%\keywords{}

%\maketitle must follow title, authors, abstract, \pacs, and \keywords
\maketitle
%\section{Introduction}
Although in most of the materials electronic spin magnetization dominates their magnetic behaviors, in a few unusual magnetic materials orbital magnetization can be significant and even dominant.~\cite{Taylor2002,Gotsis2003,Qiao2004,Laguna2010,Kolchinskaya2012}  
Though its implications on the fascinating properties of a broader range of materials \cite{Murakami2006,Xiao2006,Xiao2007,Wang2007,Essin2010} is 
self-motivating, yet much of the attention has been paid to its spin counterpart.
In contrast to atomic and molecular species where only local electronic circulation is possible, the orbital magnetization of a crystalline material comes from intracellular and intercellular circulations.~\cite{Xiao2005,Thonhauser2015,Ceresoli2006,Shi2007,Chen2011,Bianco2013}
Often concomitant to magnetism is the most intricate issue of strong correlation effect, which usually incurs dramatic changes in the electronic structure. This begets a natural question whether and in what manner electronic correlation will affect the orbital magnetization in crystalline materials.

This interesting problem has been approached with various theoretical framework in previous studies, however associated with them are their natural shortcomings. 
For example, in density-functional theory (DFT) with Hubbard $U$ correction, correlation effects were included only at static mean-field level~\cite{Nikolaev2014}. Studies on the related issue of exciton $g$-factor renormalization based on GW approximation~\cite{Wozniak2020,Deilmann2020,Xuan2020}, which although accounts for many-body effects at perturbative level, only simple formalism of orbital magnetic moments for non-interacting case was used. While the modern theory of orbital magnetization has been generalized to interacting systems and applied to model systems~\cite{Nourafkan2014,Aryasetiawan2016,Acheche2019,Sjostrand2019}, its applicability when dealing with real materials is yet to be established. 
Thus, a more generic approach to this problem would be to apply the formalism for interacting case on a real material with proper treatment of electronic correlations beyond static mean-field level.

The stabilization of long-range magnetic order in layered magnetic materials, despite the celebrated Mermin-Wagner theorem~\cite{Mermin1966}, is due to the strong magnetic anisotropy through spin-orbit coupling, 
a direct consequence of the unquenched orbital moments of magnetic ions. A large number of layered materials contains transition metal elements which are  themselves magnetic in nature and exhibit strong electronic correlation. On top of that, highly anisotropic nature of chemical bonding with strong in-plane and weaker interlayer cohesion, leads to reduction of electronic dimensionality. This in turn leads to reduced screening of Coulomb interaction, hence stronger electronic correlation. Thus, the question raised above becomes highly relevant to this class of materials.
A unique advantage of these materials is 
that the individual layers can be removed and transferred to desired substrate~\cite{Ajayan2016}, yielding a quasi-two-dimensional monolayer. The orbital magnetization therein only has the out of layer-plane component due to dimensional reduction, making room for strong magnetic anisotropy.

Two-dimensional magnetism with long range magnetic order has just been established in monolayer materials, such as  Cr$_2$Ge$_2$Te$_6$ and CrI$_3$~\cite{Wang2018,Song2018,Klein2018}. In both of these Cr compounds, crystal field-splitted lower lying $t_{2g}$ orbitals are fully filled with three electrons in majority spin channel, leaving less room for unquenched 
orbital moment. This observation makes a material with less than fully filled $t_{2g}$ orbitals all the more interesting.
Recently reported VI$_3$, which is also suggested to be a layered van der Waals magnetic material, satisfy this pre-condition.
It is found to display more complicated magnetic behavior compared to CrI$_3$~\cite{Tian2019,Kong2019,Son2019,Gati2019}.
In particular, VI$_3$ is also suggested to be a Mott insulator, and  exhibits 
larger saturated magnetization along $c$ axis than in-plane direction. More specifically, Tian $et~al.$, reported a saturated magnetic moment of 2.47$\mu_B$/V along $c$ axis, slightly larger than the expected value from spin polarization, indicating an unquenched orbital magnetic moment.
Past study on YVO$_3$ employing Hartree-Fock approach also reported small but non-zero orbital magnetization~\cite{Nikolaev2014}.
%It is expected that strong correlation effect will play an important role in understanding of the electronic and magnetic properties of VI$_3$.
Based on the above considerations, VI$_3$ seems to be a good platform to explore the effect of electronic correlation on orbital magnetization.

\begin{figure}[b]
	\centering
	\includegraphics[width=70.3mm]{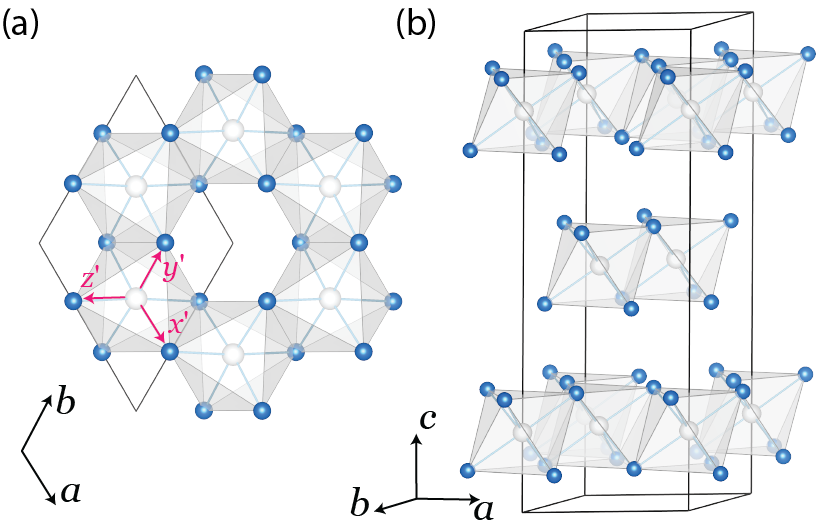} \\
	\caption{Crystal structures of (a) monolayer,  and (b) bulk $R\bar{3}$. The local Cartesian coordinates for the octahedral coordination,  $x^\prime-y^\prime-z^\prime$, are indicated with red arrows.
	}
	\label{fig:struct}
\end{figure}

In this paper, we study the effect of dynamical correlations on the magnetic properties of VI$_3$. To this end, we use a combination of %\sout{density-functional theory (DFT)} 
DFT+$U$~\cite{wien2k1, wien2k2} and self-consistent DFT + dynamical mean-field theory~\cite{Haule2010} in our study. Hereafter, we shall refer to the latter method simply as DMFT.
%combination with the density-functional theory, we report an investigation on how dynamical correlation modifies the magnetic structure of  VI$_3$. 
We find that compared to the static mean-field results from DFT+$U$ method, the dynamical correlations involved in the self-consistent DMFT framework  enhance the orbital magnetization in VI$_3$ in both the states, the high temperature paramagnetic as well as the low temperature ferromagnetic state. We find that dynamical correlation also stabilizes the orbital magnetization in the monolayer limit.
To the best of our knowledge, this would be the first study to show the role of dynamical correlation in evaluating the orbital magnetization in real magnetic materials.

VI$_3$ has been found to crystallize in different polymorphs  owing to different layer stacking, with the space groups $R\bar{3}$, $P\bar{3}1c$, $C2/m$, and $C2/c$.~\cite{Tian2019,Kong2019,Son2019,Gati2019} The structural phase transitions in VI$_3$, albeit interesting, are very subtle.
A summary of experimentally reported crystal structures at low and room temperatures can be found in Section S1 of the Supplemental Material~\cite{SM}.
For simplicity, we will focus exclusively on $R\bar{3}$ structure for the discussion here, for subtle stacking difference is not expected to impact the orbital magnetization in any way significant. 
Within each layer, V atoms form a honeycomb lattice, and each V is caged by six I$^{-}$ that form edge-sharing octahedra, as shown in Fig.~\ref{fig:struct} (a). 
In the $R\bar 3$ structure, a hexagonal primitive cell contains 3 monolayers, and the vanadium honeycomb lattices display a rhombohedral stacking along crystallographic $c$-direction, as shown in Fig. \ref{fig:struct}(b).
In subsequent calculations, we adopt experimental lattice parameters for bulk $R\bar 3$~\cite{Tian2019}, whereas for a monolayer the lattice and ions are relaxed using the density-functional theory
(see the Supplemental Material~\cite{SM} for details of calculations).

Recent experiments show that the bulk VI$_3$ exhibits an optical bandgap of 0.6$\sim$0.7 eV~\cite{Kong2019,Son2019}, in contrast to a metallic band structure from a DFT electronic structure calculation as shown in Fig. S1 in the Supplemental Material~\cite{SM}. By including the static correlation described by Hubbard $U$ (i.e., DFT+$U$), the calculated band structures recover the insulating nature for reasonable $U$ values (see Fig. S1~\cite{SM}), indicating that VI$_3$ is indeed Mott insulating. 
We choose a value of $U_{\text{eff}}=U-J = 4$ eV by matching the computed band gaps (0.67 and 0.64 eV for  $R\bar{3}$ and $P\bar{3}1c$, respectively) to experimentally observed optical bandgap at room temperature (0.60 and 0.67 eV for  $R\bar{3}$~\cite{Kong2019} and $P\bar{3}1c$~\cite{Son2019}, respectively). From the projected band structure and density of states for $R\bar 3$ shown in Fig. S3(a-b)~\cite{SM}, it can be seen that both $t_{2g}$ and $e_g$ orbitals at V atoms hybridize with  $p$ states of iodines. This hybridization is significantly enhanced when compared to the result without $U$ and magnetism (see Fig.S4~\cite{SM}). It will shortly be seen that this enhanced mixing may lead to overestimation of the extended contribution to the orbital magnetization.

\begin{figure}
	\centering
	\includegraphics[width=66mm]{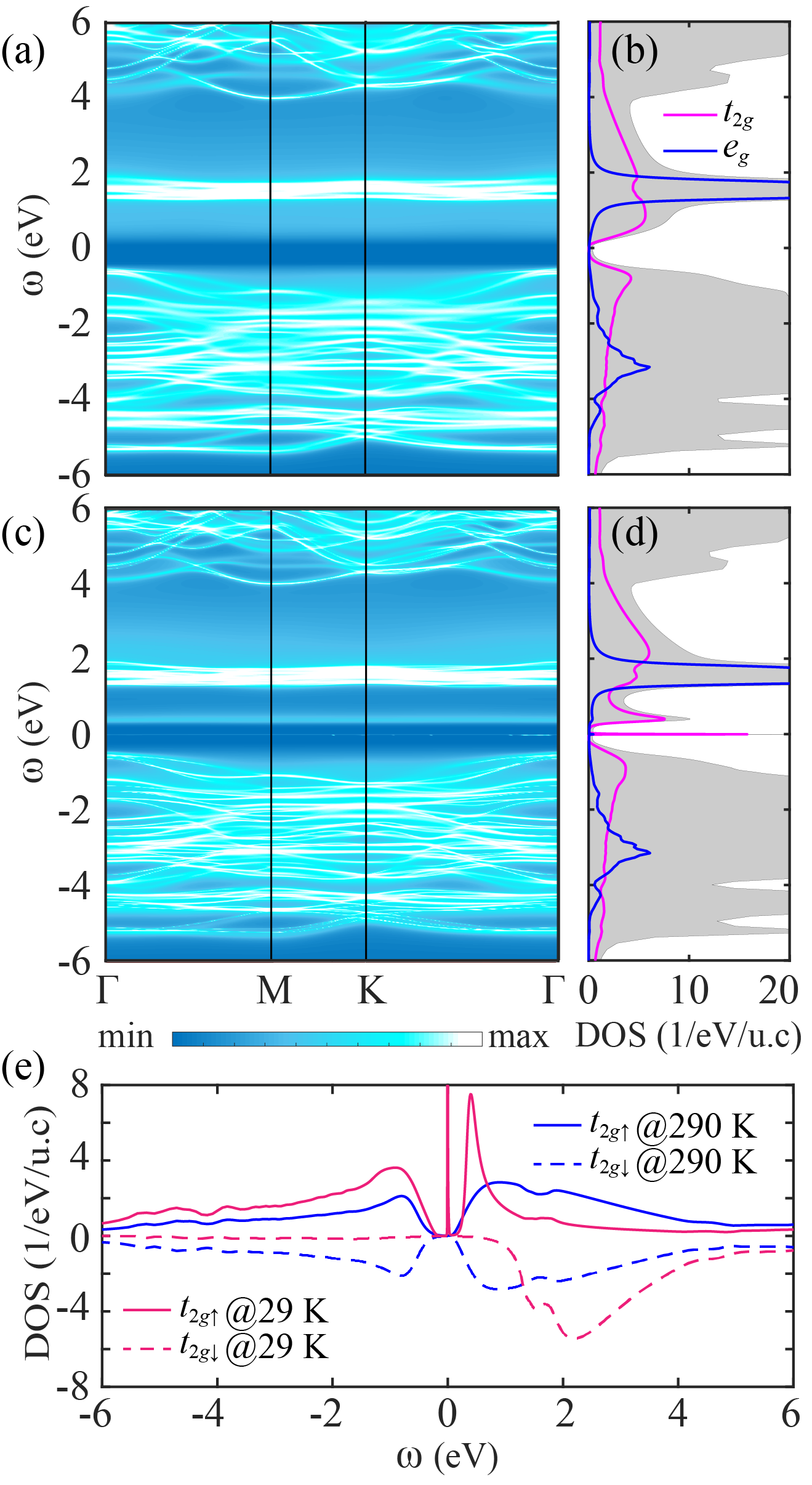} \\
	\caption{(a) and (c) are momentum resolved spectral functions for bulk VI$_3$ in the $R\bar{3}$ space group using DMFT method under 290 K and 29 K, respectively. (b) and (d) are corresponding projected density of states. (e) Spin resolved density of states.
	}
	\label{fig:dosband}
\end{figure}

As DFT+$U$ calculations only capture the static part of electronic correlation, self-consistent DMFT method is employed to incorporate the dynamical correlation~\cite{Haule2010}. The starting point of a DMFT calculation is a nonmagnetic DFT calculation without $U$, which shows a signifiant crystal field splitting between $t_{2g}$ and $e_g$ of 2 eV (see Fig.S4~\cite{SM}). Since the crystal field splitting is larger than the band widths of both $t_{2g}$ and $e_g$ sets, Hubbard $U$ is added only to the $t_{2g}$ set in DMFT calculations. 
We find that inclusion of $e_g$ orbitals into our correlated subspace does not change our main conclusions concerning the orbital magnetizations~\cite{SM}.
To obtain the spin- and momentum-resolved spectral functions in DMFT calculations, we have performed analytical continuation of self-energies using the maximum-entropy method~\cite{Haule2010}. Figure \ref{fig:dosband}(a) and \ref{fig:dosband}(c) are momentum-resolved spectral functions for high temperature (290 K) paramagnetic and low-temperature (29 K) ferromagnetic phases, respectively.
In contrast to the DFT+$U$ result, due to the dramatical renormalization from electronic correlation, the spectral weight of $t_{2g}$ becomes highly smeared out along the energy axis, although the Mott gap is clearly visible in both ferro- and paramagnetic regimes.

Experimentally, VI$_3$ is a ferromagnet with out-of-plane magnetic moments (i.e., along $z$ direction) and  Curie temperature of $T_c\sim50$~K~\cite{Tian2019,Kong2019,Son2019,Gati2019}. 
%The DMFT calculation indeed captures the phase transition.
Our DMFT calculations can reproduce both the low-temperature ferromagnetic and high-temperature paramagnetic phases.
As seen in the spin-resolved density of states shown in Fig.~\ref{fig:dosband}(e), the two spin projections of $t_{2g}$ orbitals are equally populated at 290 K, leading to the paramagnetic phase. At 29 K, the spin up component dominates and the system develops a ferromagnetic order.
In the low-temperature ferromagnetic phase, there is a quasiparticle peak around the chemical potential as shown in Fig.~\ref{fig:dosband}(d) and \ref{fig:dosband}(e). This dynamical singlet, attributable to Kondo resonance, is insensitive to the parameters used in analytical continuation, but is quickly obliterated by thermal fluctuations as the temperature rises.
In the ferromagnetic phase, the local spin moment on each V atom obtained from DMFT and DFT+$U$ is 1.88 and $2.00~\mu_{\text{B}}$, respectively, which is close to the expected $S=1$ state.
We will examine next the orbital magnetization at both DFT+$U$ and DMFT levels.

In DFT+$U$ static mean-field approximation,  the orbital magnetizations can be efficiently calculated using the non-interacting formula in the low temperature limit,
\begin{equation}
\boldsymbol{M}(\boldsymbol k)=  -\frac{\mathrm{i}e}{2\hbar }\sum_{n}   
f_{n\boldsymbol{k}}  \left\langle \frac{\partial u_{n\boldsymbol{k}}}{\partial \boldsymbol{k}} \right\vert \times (H_{\boldsymbol{k}}+\varepsilon_{n\boldsymbol{k}} -2\mu) \left\vert \frac{\partial u_{n\boldsymbol{k}}}{\partial \boldsymbol{k}}  \right\rangle, 
\label{nonint}
\end{equation}
where $u_{n\boldsymbol{k}}$ is the cell-periodic part of the Bloch function of the $n^{th}$ band at crystal momentum $\boldsymbol{k}$, $\varepsilon_{n\boldsymbol{k}}$ is the band dispersion and $\mu $ is the chemical potential. Brillouin zone summation of $\boldsymbol{M}(\boldsymbol k)$ then yields the total orbital magnetic moment (See Supplemental Material~\cite{SM} for calculation details).
The value of the orbital magnetization for bulk $R\bar 3$ phase is about 0.02 $\mu_B$ per V, accounting for about 1\% of the spin moment. The orbital magnetic moment of monolayer  VI$_3$
is computed to be essentially zero within this approach.

To further incorporate dynamical correlation in the self-consistent DMFT level, the generalized formula for orbital magnetization expressed in terms of interacting Green's functions is used~\cite{Nourafkan2014},
\begin{small}
	\begin{align}
 M^c(\boldsymbol k)=& \frac{\mathrm{i}e}{2\hbar\beta}\sum_{\omega_n}   ~ \epsilon^{abc} ~ \text{tr} \left\{ \left[H_0-\mu+\frac{\Sigma}{2}\right]G v_a(\boldsymbol{k}) G v_b(\boldsymbol{k}) G \right\}
%	 \\
%	&+\frac{1}{2N_{\boldsymbol{k}}\beta}\sum_{\boldsymbol{k}, \omega_n}  ~ \text{tr} \left\{  \left[H_0 +(\mathrm{i}\omega_n-\mu)\boldsymbol{1}\right] G  (\partial_{B_a} \Sigma^{(\boldsymbol{B})}  )_{\boldsymbol{B}=0}  G\right\} 
	\label{fig:dmftorb}
	\end{align}
\end{small}
where $a,b,c$ refer to the Cartesian axes $x,y,z$, $\omega_n=(2n+1)\pi/\beta$, $G(\boldsymbol{k},\mathrm{i}\omega_n)$ and $\Sigma(\mathrm{i}\omega_n)$ are Matsubara frequencies, interacting Green's functions, and self-energies, respectively. $v_a(\boldsymbol{k})=-\partial_{k_a} G^{-1}$ is the velocity operator.
In the above formula, a term involving the derivative of self-energy against {magnetic field} $B$ has been dropped, as the DMFT self-energy cannot depend on $\boldsymbol{B}$ linearly~\cite{Nourafkan2014}. For $\Sigma=0$ this formula will reduce to the non-interacting case given in Eq.~\ref{nonint}, which can be confirmed by both explicit derivation and numerical calculation. The Matsubara summation is converged with $|n|\le 600$ for $\beta=40~(290~\text{K})$, and $|n| \le 2000$ for $\beta=400~(29~\text{K})$.

Remarkably, in the DMFT calculations the computed orbital magnetizations are greatly enhanced compared with the DFT+$U$ results, as listed in Table.~\ref{fig:orbcmp}. The DMFT method gives consistent values of $M^z$ for all bulk structures and a monolayer. This is reasonable in the sense that the local octahedral crystal fields imposed on V atoms are nearly identical in these structures.
It should be pointed out that Eq.~\ref{fig:dmftorb} is derived without considering the entropic contribution, and only suitable for low temperatures.~\cite{Shi2007} Although,  the values of $U$ and $J_{\text{H}}$ in our DMFT calculations are chosen by matching the computed spectral gap at 290 K to experimental optical gap at room temperature~\cite{Kong2019,Son2019}, more calculations show that the calculated orbital magnetizations are similar within reasonable range of $U$ and $J_{\text{H}}$ values (see Table.~S1~\cite{SM}).

\begin{table}[H]
	\centering
	\caption{Calculated orbital magnetization along $z$-direction for bulk $R\bar{3}$, $P\bar{3}1c$, $C2/m$, $C2/c$, and monolayer (ML) VI$_3$ using different methods and temperatures as indicated in each case. Units: $\mu_{\text{B}}/$V.
	}
	\begin{tabular}{cc|p{1cm}<{\centering}|p{1cm}<{\centering}|p{1cm}<{\centering}|p{1cm}<{\centering}|p{1cm}<{\centering}}
		\hline
		\hline
		& ~~    &$R\bar{3}$& $P\bar{3}1c$ & $C2/m$ & $C2/c$ &ML	 \\
		%\hline
		\hline
		&DFT+$U$	~		&0.021   &0.026 &0.003 & 0.020&0.001\\
		\hline
		&DMFT (290 K)	&0.079   &0.084 &0.079& 0.081&0.084\\
		\hline
		& DMFT (29 K)	&0.080   &0.085 &0.079& 0.085&0.085\\
		\hline
		\hline
	\end{tabular}
	\label{fig:orbcmp}
\end{table}

\begin{figure}
	\centering
	\includegraphics[width=70mm]{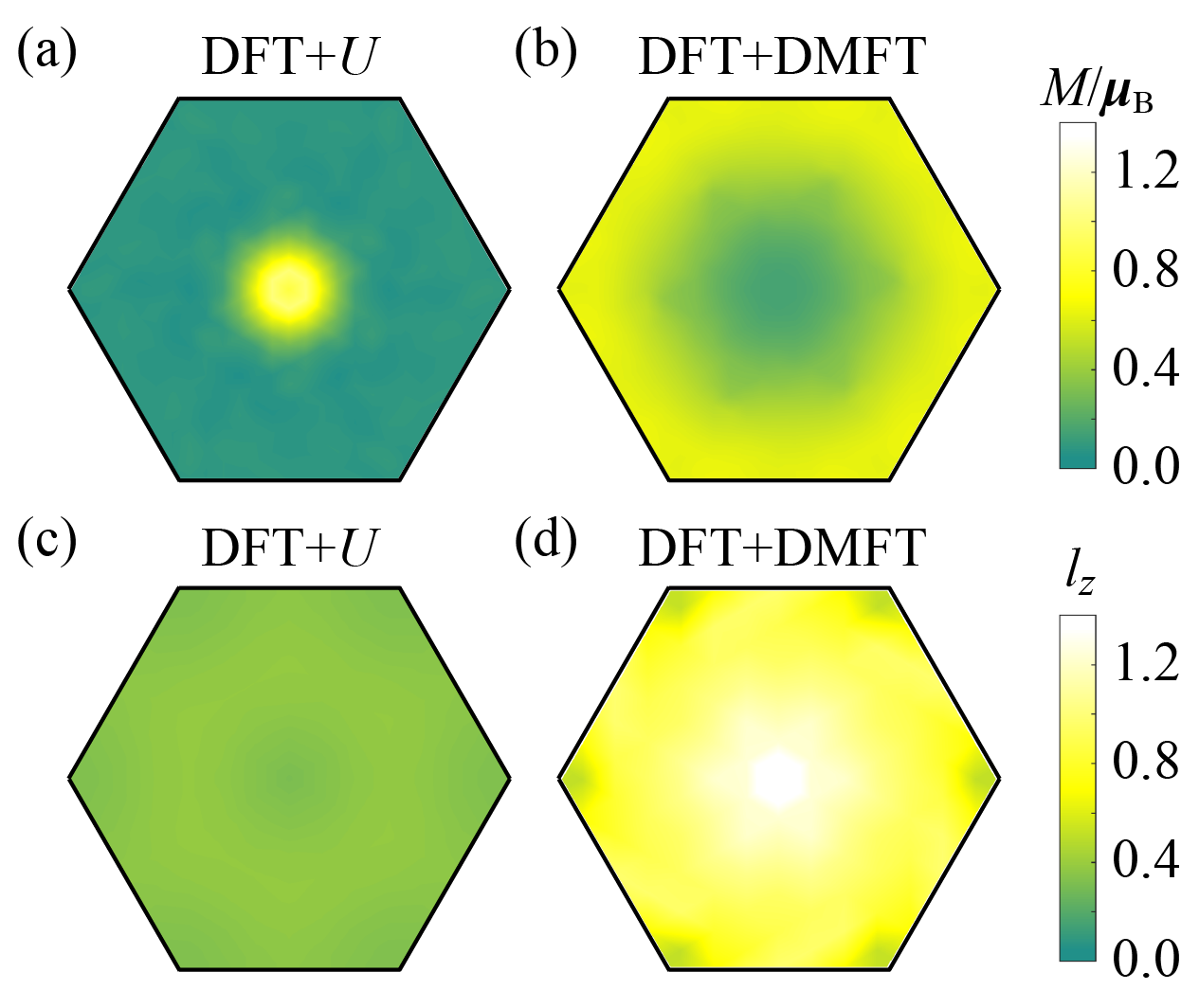} \\
	\caption{(a) and (b) are calculated orbital magnetization distribution in the Brillouin zone (at $k_z=0$ plane) for bulk $R\bar{3}$ using DFT+$U$ and DMFT (29 K), respectively. (c) and (d) are calculated orbital angular momentum distribution in the Brillouin zone (at $k_z=0$ plane) for bulk $R\bar{3}$ using DFT+$U$ and DMFT (29 K), respectively.
	}
	\label{fig:r3Mklzk}
\end{figure}

In order to further analyze how the dynamical correlation influence orbital magnetization, we plot the orbital magnetization $ M^z(\boldsymbol k)$ across the Brillouin zone from both DFT+$U$ and DMFT calculations for $R\bar{3}$ structure,  as shown in Fig.~\ref{fig:r3Mklzk}(a) and (b). It is obvious that the distributions of orbital magnetization in the Brillouin zone are quite different for these two methods. In the DFT+$U$ method, the main contribution to orbital magnetization comes from the Brillouin zone center, while the orbital magnetization accumulates mainly along the Brillouin zone boundaries for the DMFT result. The distribution of $ M^z(\boldsymbol k)$ from DFT+$U$ is more localized compared to the DMFT result, which means the orbital magnetization in real space should be more extended in the DFT+$U$ result. This difference can be partially attributed to the enhanced $d$-$p$ mixing in the DFT+$U$ electronic structures as mentioned earlier. Another source of the difference is the dramatical renormalization of the electronic structures through 
%real part of 
the DMFT self-energy, as shown in the spectral functions above (Fig. \ref{fig:dosband}(a) and \ref{fig:dosband}(c)). The overall effects lead to an over four-fold enhancement of total orbital magnetization compared to DFT+$U$ result. It is noteworthy that the orbital magnetization is along the $z$ direction, thus it will lead to strong anisotropy of saturation magnetization, which is consistent with the experimental observation~\cite{Tian2019,Kong2019,Son2019}.

Since in atomic limit, the orbital magnetization originates from orbital angular momentum, 
we have done further analysis to elucidate the correlation effect on the orbital angular momentum of Bloch wave functions. As the net residual orbital angular momentum can come from the imbalanced occupation of $d$ orbitals of vanadium atoms, 
we have projected the Bloch wave functions to the 5 channels of the spherical harmonics $Y_2^m$ with $m=0,\pm1,\pm2$. 
%In order to obtain the orbital angular momentum along the global $z$-direction, the $d$ orbitals should be defined under global Cartesian coordinates. 
The $\boldsymbol{k}$-resolved orbital angular momentum is defined as
\begin{equation}
l_z(\boldsymbol{k})=\sum_{n\in \text{occ}}\sum_{m}m|\langle\psi_{n\boldsymbol{k}}|Y^m_2\rangle|^2,
\end{equation}
where the summation is over the occupied states. $\psi_{n\boldsymbol{k}}$ is the Bloch wave functions. The results for bulk $R\bar{3}$ structure are shown in Fig.~\ref{fig:r3Mklzk} (c) and (d). 
It can be seen that for both the methods, distributions of orbital angular momentum in $\boldsymbol{k}$-space are almost uniform, especially for the DFT+$U$ method. The reason behind such a behavior can be understood in terms of strong localization of $d$ orbitals on V atoms, leading to weak $\boldsymbol{k}$-dependence of orbital angular momentum. However, the values of orbital angular momentum from DMFT is greatly enhanced compared to that from DFT+$U$. The integrated $l_z$ for DFT+$U$ and DMFT are 0.060/V and 0.158/V, respectively. 
Note that for an atomic model with 2 electrons filling the $t_{2g}$ orbitals of V atoms, according to the Hund’s rules, the orbital magnetic moment should be 1$\mu_B$/V and antiparallel to the $2\mu_B$/V spin moment~\cite{Georgios2018}. However, the experimental observations show that the out-of-plane magnetic moments are larger than the in-plane ones ~\cite{Tian2019,Kong2019,Son2019}, indicating that the orbital magnetic moment should be parallel to the spin moment. In our DMFT study, we find the orbital moments of $\sim0.08\mu_B$/V 
parallel to spin moments, consistent with the experimental observations. Our results imply that the orbital magnetization in real materials cannot be simply explained by the atomic model.

Hence, a unified physical picture for dynamically enhanced orbital magnetization in VI$_3$ emerges from the above comparison. Despite a stronger $p$-$d$ mixing in DFT+$U$ that increases the intercellular electronic circulations of electrons (corresponding to the hot spot in Fig.~\ref{fig:r3Mklzk} (a)), it  suppresses the (potentially more important) intracellular circulations at the same time. The overall effect is minimal occupational imbalance among $l_z$ channels, and quenched orbital angular momentum.
In the self-consistent DMFT framework, the localization behavior of correlated electrons in VI$_3$ is found to increase the occupational imbalance of $l_z$ channels dramatically. Although the orbital hybridizations in crystals will still suppress the orbital magnetization, the greatly enhanced residual orbital angular momentum eventually leads to a significantly enhanced orbital magnetization in DMFT approach when compared to that from DFT+$U$.

Futher insights can be obtained by separating the orbital magnetization into local ($\boldsymbol{M}_{\text{LC}}$) and itinerant ($\boldsymbol{M}_{\text{IC}}$) parts, leading support to the foregoing picture. As defined by Thonhauser $el~al.$~\cite{Thonhauser2015},
$\boldsymbol{M}_{\text{LC}}= -\mathrm{i}e/(2N\hbar) \sum_{n\boldsymbol{k}}   
f_{n\boldsymbol{k}}  \left\langle \nabla_{\boldsymbol{k}} u_{n\boldsymbol{k}}\right\vert \times H_{\boldsymbol{k}} \left\vert \nabla_{\boldsymbol{k}} u_{n\boldsymbol{k}}  \right\rangle$,
and 
$\boldsymbol{M}_{\text{IC}}= -\mathrm{i}e/(2N\hbar) \sum_{n\boldsymbol{k}}   
f_{n\boldsymbol{k}}  \left\langle \nabla_{\boldsymbol{k}} u_{n\boldsymbol{k}} \right\vert \times \left\vert \nabla_{\boldsymbol{k}} u_{n\boldsymbol{k}} \right\rangle\varepsilon_{n \boldsymbol{k}}$.
%\begin{equation}
%\boldsymbol{M}^{\text{LC}}(\boldsymbol{k})=  -\frac{\mathrm{i}e}{2\hbar }\sum_{n}   
%f_{n\boldsymbol{k}}  \left\langle \frac{\partial u_{n\boldsymbol{k}}}{\partial \boldsymbol{k}} \right\vert \times H_{\boldsymbol{k}} \left\vert \frac{\partial u_{n\boldsymbol{k}}}{\partial \boldsymbol{k}}  \right\rangle, 
%\end{equation}
%and
%\begin{equation}
%\boldsymbol{M}^{\text{IC}}(\boldsymbol{k})=  -\frac{\mathrm{i}e}{2\hbar }\sum_{n}   
%f_{n\boldsymbol{k}}  \left\langle \frac{\partial u_{n\boldsymbol{k}}}{\partial \boldsymbol{k}} \right\vert \times \left\vert \frac{\partial u_{n\boldsymbol{k}}}{\partial \boldsymbol{k}}  \right\rangle\varepsilon_{n \boldsymbol{k}}.
%\end{equation}
For DFT+$U$ case, these two parts can be calculated according to the above formulas separately.
In order to compare the results given by DFT+$U$ and DMFT on the equal footing, we need some approximations for the DMFT case. 
Specifically, we used the eigen energies from the DFT step but corrected by the real part of self-energy at infinity frequency, corresponding to a shift of chemical potential, to ensure a proper electron number, i.e., $\varepsilon_{n \boldsymbol{k}}^{\text{DMFT}}\approx\varepsilon_{n \boldsymbol{k}}^{\text{DFT}}+\text{Re}\Sigma_{\infty}$.
%\begin{equation}
%\varepsilon_{n \boldsymbol{k}}^{\text{DMFT}}\approx\varepsilon_{n \boldsymbol{k}}^{\text{DFT}}+\text{Re}\Sigma_{\infty}.
%\end{equation}
%Since our DMFT calculations are charge self-consistent, it is expected that the electronic correlation effects have been partially encoded into the DFT eigen energies and velocity matrix elements during the iterations. As a result, the calculated $M^{\text{LC}}+M^{\text{LC}}$ should be close to the total orbital magnetization, which is indeed true from our results.
As the $d$-$p$ mixing is directly related to the itinerant motion of electrons, it is indeed found that in the DFT+$U$ results, the itinerant contribution dominates the orbital magnetization ($M^z_{\text{LC}}=-0.012\mu_B/$V and $M^z_{\text{IC}}=0.033\mu_B/$V). However, the orbital magnetization is dominated by the local part in the DMFT framework ($M^z_{\text{LC}}=0.208\mu_B/$V and $M^z_{\text{IC}}=-0.148\mu_B/$V), due to the localization behavior of V atoms.

In summary, we have studied the electronic correlation effect on orbital magnetization, taking the layered van der Waals magnetic materials VI$_3$ as an example. Our calculations reveal that the dynamical correlation is crucial for evaluating the orbital magnetization in correlated materials like VI$_3$. 
The static mean-field theory based density-functional theory (in the DFT+$U$ level) is insufficient to capture the renormalization to spectral functions and the localization behavior of correlated electrons. As a result, it will underestimate the intracellular circulations contribution to orbital magnetization for these materials. 
Making use of the state-of-the-art dynamical mean-field theory, we are able to recover the dynamical correlation and give a better description to orbital magnetization, which is consistent with the experimental observation~\cite{Tian2019,Kong2019,Son2019}.
It is interesting that
the correlation effect can even stabilize the orbital magnetization to monolayer limit, which calls for further experiments to verify. Our study may inspire the research of low dimensional magnetism as well as potential spintronics applications. 
From the future perspective, effect of short-range or nonlocal correlations on the orbital magnetization would be interesting to study. It can be done using more accurate but computationally costly cluster methods~\cite{Hettler1998,Kotliar2001}, or GW+DMFT~\cite{Nilsson2018}.

\begin{acknowledgements} This work was supported by the National Natural Science Foundation of China (Grant No. 11725415 and No. 11934001), the Ministry of Science and Technology of  China (Grant No. 2018YFA0305601 and No. 2016YFA0301004), and by the Strategic Priority Research Program of Chinese Academy of Science (Grant No. XDB28000000).
\end{acknowledgements}

%\bibliography{library}

\end{document}